\documentclass[12pt]{article}
\usepackage{epsfig}
\begin{document}
\renewcommand{\arraystretch}{1.5}

\rightline{UG-12/97}
\rightline{EHU-FT/9708}
\rightline{hep-th/9801082}
\vspace{3truecm}
\centerline{\bf UNSTABLE VORTICES DO NOT CONFINE}
\vspace{2cm}
\centerline{A.~Ach\'ucarro${}^{1,2}$, M.~de Roo${}^1$ and 
  L.~Huiszoon${}^1$}
\vspace{1.0truecm}
\centerline{${}^1$\it Institute for Theoretical Physics}
\centerline{\it Nijenborgh 4, 9747 AG Groningen}
\centerline{\it The Netherlands}
\vspace{0.5truecm}
\centerline{${}^2$\it Department of Theoretical Physics}
\centerline{\it The University of the Basque Country UPV-EHU}
\centerline{\it Apt 644, 48080 Bilbao}
\centerline{\it Spain}
\vspace{2truecm}
\centerline{ABSTRACT}
\vspace{.5truecm}

Recently, a geometric model for the confinement of magnetic charges in
 the context of type II string compactifications was constructed by
 Greene, Morrison and Vafa \cite{GMV96}. This model assumes the
 existence of stable magnetic vortices with quantized flux in the low
 energy theory.  However, quantization of flux alone does not imply
 that the vortex is stable, since the flux may not be confined to a
 tube of definite size.  We show that in the field theoretical model
 which underlies the geometric model of confinement, static,
 cylindrically symmetric magnetic vortices do not exist. While our
 results do not preclude the existence of confinement in a different
 low-energy regime of string theory, they show that confinement is not
 a universal outcome of the string picture, and its origin in the low
 energy theory remains to be understood.

\newpage

\noindent{\bf 1.\ Introduction}
\vspace{0.5truecm}

In a recent paper a mechanism for confinement of magnetic flux was put
forward in type II string compactifications on Calabi-Yau manifolds,
where magnetic states can arise from D branes wrapped around
non-trivial chains \cite{GMV96}.  In this picture, chains must be
attached to other chains in order to make a closed 3-cycle on which to
wrap the brane. In four dimensions, the configuration would look like
pairs of oppositely charged magnetic monopoles joined by flux tubes
(sometimes known as ``dumbells''), thus providing a mechanism
for confinement.

 From the field theory point of view, the low energy theory contains
sixteen hypermultiplets charged under fifteen U(1) gauge groups, in
such a way that the condition for finite energy per unit length of a
vortex-like configuration translates into a correlation between the
various windings at infinity which ultimately leads to the
quantization of magnetic flux. Taken together with the fact that the
Higgs mechanism is operating in this model one would be tempted to
conclude that there are stable vortices with a width given by the
inverse vector mass, similar to those found in the Abelian Higgs model
\cite{A57, NO73}.

It is the purpose of this letter to show that this assumption may not
be justified. We will see that the low energy theory discussed in
\cite{GMV96} does not admit stable static axisymmetric magnetic
vortices of a fixed width.  Any such configuration immediately decays
by expanding its core radius indefinitely. While our result does not
preclude the existence of non-axisymmetric flux tubes, it is very
unlikely that such stable structures exist.  The physical origin of
the instability can be traced back to the repulsion of magnetic field
lines which, unlike for the Abelian Higgs model, is not compensated
here by an increase in potential energy during the expansion because
the potential has flat directions.  It is difficult to see how
non-axisymmetric configurations could circumvent this problem.

It is well known that the quantization of magnetic flux does not
guarantee its confinement into flux tubes, even when the gauge bosons
are made massive by the Higgs mechanism. Consider, for instance, the
so-called semilocal strings \cite{VA91, H92} which arise in the
Weinberg-Salam model in the limit of zero $SU(2)$ coupling, a model
that has several features in common with the one analysed here. In
particular, the relative shortage of gauge field degrees of freedom
also introduces a correlation between the winding of the scalars at
infinity. As a result, magnetic flux is quantized, with the various
sectors separated by infinite energy barriers. However, it has been
shown that the stability of vortex solutions in this model depends on
the ratio of the scalar and vector masses. If $m_{scalar} <
m_{vector}$ there are stable vortices whose width is related to the
inverse vector mass (they are in fact identical to the Nielsen-Olesen
vortices in the Abelian Higgs model).  However, when $m_{scalar} >
m_{vector}$, there are no stable vortices; the magnetic core tends to
expand indefinitely while conserving magnetic flux. Thus, magnetic
flux is quantized but not confined in this case.  (This is not so
surprising; the quantization of magnetic flux has to do with the
behaviour of the fields far away from the vortex, whereas stability
depends on their behaviour at the core, and unless there is a
topological reason to link these two distance scales, they will be
independent of each other).  Moreover, when $m_{scalar} = m_{vector}$,
the vortices saturate a Bogomol'nyi bound (which automatically ensures
their stability) but this is still not enough to guarantee the
confinement of magnetic flux to tubes of a definite size (see
\cite{H92, GORS92}), because there is an entire family of solutions
with the same energy and different core sizes.

What this example illustrates is that, in general, identifying a
topological invariant such as a conserved magnetic flux is not enough
to guarantee the existence of stable solutions carrying this
topological charge. Experience shows that this is a particularly
dangerous assumption when the theory contains both global and local
symmetries linked in a non-trivial way \cite{P92}. In the semilocal
model of the previous paragraph, the existence of such solutions
depends on the parameters of the theory (the scalar and vector
masses); in the model analysed here, the situation is even more
dramatic, since it seems that there are {\it no} values of the
parameters for which stable vortices exist, even in the lowest
non-zero energy sector.

Thus, as long as the string theory regime is such that its low energy
behaviour is the one discussed in \cite{GMV96}, confinement of
magnetic charges is very unlikely. Needless to say, this does not
preclude the existence of confinement in other low-energy regimes; it
merely shows that confinement is not a {\it necessary} outcome of the
string picture and remains to be understood.

In \cite{GMV96}, the existence of these confining flux tubes was argued in
the simpler example of a field theory with two hypermultiplets with
opposite charges under a single U(1) gauge field. Since the physics in
the discussion is essentially the same as in the sixteen multiplet
case, we will also consider this simplified model. In order to study
vortex solutions we will impose translational symmetry in the z
direction and reduce the model to 2+1 dimensions. We will then prove
the non-existence of static axisymmetric vortices, and discuss the
implications for the 3+1 dimensional ``dumbell'' configurations and
confinement.

\vspace{0.5truecm}
\noindent{\bf 2.\ The model}
\vspace{0.5truecm}

The simplified field theory model of \cite{GMV96} contains two $N=2$ 
hypermultiplets, each containing two physical and two
 auxiliary scalar fields (all complex) and a Dirac spinor,
 coupled to the $N=2$ Abelian vector multiplet.
 The Lagrangian (in Wess-Zumino gauge) reads~\cite{S85}
\begin{equation} \label{eq:gaugedlag}
   \mathcal{L} = \mathcal{L}_{gauge} + \mathcal{L}_{matter} + 
   \mathcal{L}_{interaction} \,,
 \end{equation}
where (implicit summation over $a$, which counts the hypermultiplets)
\begin{eqnarray}
\mathcal{L}_{gauge} & = & 
  \frac{1}{2}\,(\partial_{\mu} M)^2 + \frac{1}{2}\,(\partial_{\mu} N)^2 + 
  \frac{i}{2}\, \bar{\lambda}_i\, \gamma^{\mu}\, \partial_{\mu} \lambda^i 
  - \frac{1}{4}\, F^{\mu \nu} F_{\mu \nu} 
  + \frac{1}{2}\, \vec{D}^{\,2}\,, \nonumber \\ 
\mathcal{L}_{matter} & = & \frac{1}{2}\, D^{\mu} h_a^{\;\;i}\, 
  D_{\mu} h_{ai} + i\, \bar{\psi}_{a}\, \gamma^{\mu}\, D_{\mu}\, 
 \psi_{a} + F_a^{\;\;i}\, F_{ai}\,, \nonumber \\
\mathcal{L}_{interaction} & = & i\, q_a\, h_a^{\;\;i}\, 
   \bar{\lambda}_i\, \psi_a - i\, q_a\, \bar{\psi}_a\, \lambda^i\, h_{ai}  
  - q_a\,\bar{\psi}_a\,(M - \gamma^5\, N)\, \psi_a - \nonumber  \\
&  & \frac{1}{2}\, h_a^{\;\;i}\,(M^2 + N^2)\, h_{ai} + \frac{1}{2}\, 
       q_a\, h_a^{\;\;i}\, \vec{\tau}_i^{\;\;j}\, \vec{D}\, h_{aj}\,.
\end{eqnarray}
The $q_a$ are the charges of the hypermultiplets, so that
\begin{equation}
   D_{\mu}h_{ai} = (\partial_{\mu} + i\, q_a\, A_{\mu})\,h_{ai}\,.
\end{equation}
The supersymmetry transformations of the fields of the
 hypermultiplets take on the form
\begin{eqnarray}
\lefteqn{\delta h_{ai}} & \ \ \ = & 
  2\, \bar{\epsilon}_i\, \psi_a \,,\nonumber \\
\lefteqn{\delta \psi_a} & \ \ \ = & 
  -i\, \epsilon^{\,i}\, F_{ai} - (i\,\gamma^{\mu}\, D_{\mu} 
  + M + \gamma^5\, N)\, \epsilon^{\,i}\, h_{ai}\,, 
 \nonumber \\
\lefteqn{\delta F_{ai}} & \ \ \ =  & 
   2\, \bar{\epsilon}_i\, (\gamma^{\mu}\,D_{\mu}+i\,M-i\,\gamma_5\,N) \psi_a -2\bar{\epsilon}_j\,\lambda^j\,h_{ai},
\end{eqnarray}
while the fields of the gauge multiplet transform
 as
\begin{eqnarray} 
\label{eq:gauge2trans}
\lefteqn{\delta A_{\mu}} & \ \ \ = & 
 i\,\bar{\epsilon}_i\,\gamma_{\mu}\,\lambda^i \,,
 \nonumber \\
\lefteqn{\delta M}       & \ \ \ = & 
 i\,\bar{\epsilon}_i\,\lambda^i \,,
 \nonumber \\
\lefteqn{\delta N}       & \ \ \ = & 
  i\,\bar{\epsilon}_i\,\gamma^5\,\lambda^i\,, 
 \nonumber \\
\lefteqn{\delta \lambda^i} & \ \ \ = & 
  -\frac{i}{2}\,\sigma^{\mu \nu}\,\epsilon^{\,i} F_{\mu \nu} -  
  \gamma^{\mu}\,\partial_{\mu}\,(M + 
  \gamma^5\,N)\,\epsilon^{\,i} - 
  i\,\epsilon^{\,j}\,\vec{\tau}_{j}^{\;\;i}\,\vec{D}\,, 
 \nonumber \\
\lefteqn{\delta \vec{D}} & \ \ \ = & 
  \epsilon_i\,\vec{\tau}_{j}^{\;\;i}\,\gamma^{\mu}\,\partial_{\mu} \lambda^j\,.
\end{eqnarray}

Furthermore, the Lagrangian has a global $SU(2)$ symmetry that
 rotates the two scalar fields of the multiplets. Note that there
 is no continuous symmetry that mixes the hypermultiplets, unless they
 have the same $U(1)$-charge. 

We can eliminate the auxiliary fields and get a self-interaction term
 for the scalars $h_{ai}$. The equations of motion for the
 auxiliary fields are
\begin{equation} 
  F_{ai}  =  0 \,,\qquad
 \vec{D}  =  H^{\;\;i}_j\,\vec{\tau}^{\;\;j}_i\,, 
\end{equation} 
where $H^{\;\;i}_j = - \frac{1}{2}\,q_a\,h_a^{\;\;i}\,h_{aj}$  
 ($h^{\;\;i}_a=h^*_{ai}$). When we substitute these equations back into
the Lagrangian, a term $-V(h_{ai})$ arises, with
\begin{eqnarray} 
 V(h_{ai}) \mbox{ } &=& 
    \frac{1}{2}\,\vec{D}^{\,2} \nonumber \\ 
  \label{eq:potential} &=& \frac{1}{2}\,[\,(H^{\;\;1}_2 + H^{\;\;2}_1)^2
 + (i\,H^{\;\;1}_2 - i\,H^{\;\;2}_1)^2 + (H^{\;\;1}_1 -
 H^{\;\;2}_2)^2\,] \,.
\end{eqnarray} 
Note that $(H^{\;\;i}_j)^* =
 H^{\;j}_i$, so the potential is a sum of three positive terms. 
 In what follows we will limit ourselves to the case where the 
 two hypermultiplets have $U(1)$-charges $q_a=(1,-1)$, as is done 
 in~\cite{GMV96}.
 The minimum of the potential, $V(h_{ai}) = 0$, is
 obtained for 
\begin{eqnarray} 
 H^{\;\;1}_2 = & - {1 \over 2}\,[\, h_{11}^* h_{12} 
   - h_{21}^* h_{22}\,] & = 0  \,,\nonumber \\ 
\label{eq:minimumb}
 H^{\;\;1}_2 = & - {1 \over 2}\,[\, h_{12}^*
   h_{11} - h_{22}^* h_{21} \,] & =  0 \,, \\ 
 H^{\;\;1}_1 - H^{\;\;2}_2 = & 
  -{1\over 2}\, [\,|h_{11}|^2 +|h_{22}|^2 - |h_{12}|^2 -
  |h_{21}|^2\, ] & = 0 \,.\nonumber
\end{eqnarray}

In order to find the vacuum manifold, we follow \cite{GMV96} and 
 parametrize the complex scalar fields as:
\begin{equation}
 h_{ai} = r_{ai}\, e^{\,i\,\theta_{ai}} \,.
\end{equation}
Equations~(\ref{eq:minimumb}) now become
\begin{eqnarray} 
 e^{\,i\,(\theta_{11} - \theta_{12})} & = &
 e^{\,i\,(\theta_{21} - \theta_{22})}\,, \nonumber \\ 
 r_{11}\,r_{12} & =
 & r_{21}\,r_{22}\,, \nonumber \\ 
 (r_{11})^2 - (r_{21})^2 & = &
 (r_{12})^2 - (r_{22})^2 \,.
\end{eqnarray} 
These equations are solved (up
 to a factor $2 k\pi$ in the angles) by
\begin{eqnarray} 
\theta_{11} - \theta_{12} & = & \theta_{21} - \theta_{22} \,,
 \label{eq:theta} \\
 r_{1i} & = & r_{2i} \label{eq:r} \,.
\end{eqnarray} 

Consider the bosonic sector. The contribution to the energy from the
 scalars $M$ and $N$ is positive definite, and they must tend to zero
 at infinity, so we set $M=N=0$. The model then contains four complex
 scalars and one $U(1)$ gauge field. 
 Since we are interested in the possible existence of magnetic vortex
 solutions, we will now reduce the problem to 2+1 dimensions by
 imposing translational symmetry in the z-direction, and consider
 static solutions. Specifically, we require all fields to be independent of
 $t$ and $z$ (note that, in principle, the scalar fields could have a
 non-trivial dependence on these coordinates which would lead to
 spinning or electrically charged configurations, but they all have
 higher energy). For the same reason we take $A_t = A_z = 0$. The
 electric field is zero, and the only component of the magnetic field,
 $B$, is in the $z$-direction.

The energy per unit length becomes ($m,n=1,2$):
 \begin{equation} \label{eq:energyfunctional}
	{\cal E} = \int d^2\!x\,[\,\frac{1}{2}\,|D_m h_{ai}\,|^2 + \frac{1}{4}\,F_{mn}^{\;2} + V(h_{ai})\,]
 \end{equation}
where $V(h_{ai})$ is given in (\ref{eq:potential}).
In order to have a finite energy string, we require 
\begin{equation}
\label{eq:condition} 
	D_m h_{ai}  \to  0 \,, \qquad
	F_{mn}  \to   0 \,, \qquad
	V(h_{ai})  \to   0 \,,
\end{equation}
as $r \rightarrow \infty$ which, together with (\ref{eq:theta}) and
(\ref{eq:r}),  lead to the following asymptotic behaviour for the
fields: 
\begin{eqnarray} 
h_1 & \equiv & \pmatrix{h_{11} \cr h_{12} \cr} \to   \pmatrix{c_1 \cr c_2 \cr} e^{-i\,n\,\theta} \,,
 \nonumber \\
h_2 & \equiv & \pmatrix{h_{21} \cr h_{22} \cr}  \to   \pmatrix{c_1 \cr c_2 \cr} e^{\,i\,\Delta} e^{\,i\,n\,\theta}\,,
 \nonumber \\
A_{\theta} & \to  & {n \over r} \;,\;\;\; A_r \to  0
\label{eq:solution} 
\end{eqnarray} 
for $r \rightarrow \infty$. Here $c_i$ are arbitrary complex constants,
 $\Delta$ is real.

These boundary conditions are analogous to those of Nielsen-Olesen and
 semilocal vortices in that they 
 correspond to a winding of the hypermultiplets
 around a gauge orbit at infinity ($n$ is the winding number of the
 configuration). Due to the relative shortage of degrees of freedom,
 the windings are correlated, not only within each hypermultiplet but
 also between $h_1$ and $h_2$.  Note that when the winding numbers are
 not correlated like this, the energy diverges, so the various winding
 sectors are separated by infinite energy barriers. Indeed, magnetic
 flux is quantized,
\begin{equation} \label{eq:totalflux}
\Phi = \int d^2 x B = \oint r\,d\theta\,A_{\theta} = 2\,\pi\,n\,,
\end{equation} 
but we will see in the next section that, unlike for
the Abelian Higgs and its semilocal extensions, there are no cylindrically
symmetric vortex solutions in this case.

\vspace{0.5truecm}
\noindent{\bf 3.\ Static, axisymmetric configurations}
\vspace{0.5truecm}

In what follows we will restrict ourselves to $n=1$, but our results
generalize trivially to any winding number.

The condition of axial symmetry means that, when the solution is
rotated around the $z$-axis, the rotated configuration is related by
symmetry to the original one. Consider first the gauge fields. Under
infinitesimal rotations one should find $ \partial_\theta A_m (r,\theta)
= \partial_m \alpha(r,\theta)$.  We can analyse these conditions by
choosing the gauge $A_r = 0$, in which case the function $\alpha$ is
restricted to be independent of $r$, and therefore $A_\theta = \alpha
(\theta)/r + v(r)$ (where $v(r)$ is an arbitrary function of $r$).  We can
gauge-transform to $ A_\theta = v(r)$, which determines the function
$\alpha$ up to an arbitrary constant. 
The residual invariance is a global $U(1)$.
Now we turn to the scalar fields. It is straightforward to show 
that  the most
general configuration for the scalar hypermultiplets compatible 
with cylindrical symmetry is 
\begin{equation} h_1 = \left ( {g_1 (r) } \atop {f_1(r) } \right
	)\,e^{-i\,\theta} \mbox{ , } h_2 = \left ( {g_2(r) } \atop {f_2 (r)}
	\right )\, e^{\,i\,\Delta }e^{\,i\,\theta} \,,
\end{equation} 
with boundary conditions 
$f_a,g_a \rightarrow 0$ as $r \rightarrow 0$ and $g_a
\rightarrow c_1$, $f_a \rightarrow c_2$ as $r \to \infty$.

We now prove that there are no stable axisymmetric solutions to the
equations of motion by showing that the energy of such a configuration
can always be lowered by a continuous deformation which respects the
boundary conditions. Indeed, the energy of the family
\begin{eqnarray} 
h_1(\xi) & \equiv & \left ( \begin{array}{c}
 (1-\xi)\,g_1 + \xi\,g_2 \\ (1-\xi)\,f_1 + \xi\,f_2 \end{array} \right
 ) \,e^{-i\,\theta}\,, \nonumber \\
 h_2(\xi) & \equiv & \left ( \begin{array}{c}
 \xi\,g_1 + (1-\xi)\,g_2 \\ \xi\,f_1 + (1-\xi)\,f_2 \end{array} \right
 ) \,e^{\,i\,\Delta}e^{\,i\,\theta}\,,  \nonumber 
\end{eqnarray} 
(with the same $A_\mu$) is given by
 ${\cal E}_g (\xi) +  {\cal E}_p (\xi) + {\cal E}_m (\xi)$, where
\begin{eqnarray} 
 {\cal E}_g (\xi) &=&  {\cal E}_g (0)  + \xi (\xi-1) {\cal A}\,, \nonumber\\ 
 {\cal E}_p (\xi) &=&  (1 - 2 \xi)^2 {\cal E}_p (0)\,, \\
 {\cal E}_m (\xi) &=&  {\cal E}_m (0)\,, \nonumber
\end{eqnarray}
are the energy contributions from the scalar gradients, scalar
 potential and magnetic field respectively, and $\cal A$ is some
 positive constant. The energy decreases monotonically from
 $\xi = 0$ (our starting configuration) until it reaches a minimum at
 $\xi = 1/2$. Moreover, at $\xi = 1/2$ the potential energy is zero and 
 therefore the energy can be lowered even further by letting the
 magnetic core expand. Consider the family of expanding configurations 
 for fixed magnetic flux
  ($\hat h_{a} \equiv h_{a}(\xi = 1/2)$,
 $\hat{\cal E} \equiv {\cal E}(\xi=1/2)$)
\begin{eqnarray}
 h_{a, \lambda} (r, \theta) & \equiv & {\hat h}_a ( {r\over \lambda}, 
 \theta) \,,\\
 A_{m, \lambda}(r, \theta) & \equiv & {1\over \lambda} {A}_m({r\over
\lambda}, \theta)\,,
\end{eqnarray}
whose energy per unit length is 
\begin{equation}
{\cal E}_\lambda  = {\hat {\cal E}}_{g} + {{\hat {\cal E}}_{m} \over
{\lambda^2}} \,. \\
\end{equation}
The energy decreases monotonically as $\lambda \to \infty$ (this is of
course a variation of the scaling argument used to prove
Derrick's theorem ~\cite{D64}).  This implies that, for a cylindrically
symmetric configuration, the magnetic flux can never be confined. The
magnetic field lines tend to spread to infinity, since this lowers the
energy. 

It is straightforward to check that the instability would still be
there if we had allowed for non-zero scalars $M$ and $N$.
Finally, it should be obvious that our results apply unchanged to the
low energy approximation of the model discussed in \cite{GMV96}, a
model containing sixteen hypermultiplets charged under fifteen
$U(1)$'s.

\vspace{0.5truecm}
\noindent{\bf 4.\ Discussion}
\vspace{0.5truecm}

We have argued that stable, infinitely long magnetic vortices almost
certainly do not exist in this supersymmetric model.  While our proof
only concerns axisymmetric vortices, it is hard to see how
non-axisymmetric solutions could avoid the instability described in
the previous section.  It should be stressed that the total magnetic
flux in each sector is topologically conserved. Thus, in general,
identifying a topological invariant of the theory is not enough to
guarantee the existence of stable solutions carrying this topological
charge, particularly if the theory contains both global and local
symmetries \cite{P92}, as is the case in many supersymmetric
models. Finding a higher dimensional origin for such a topological
invariant, while mathematically very compelling, may be of limited
assistance in understanding the low energy spectrum of the
compactified theory.

If we now consider 3+1 dimensional ``dumbell'' configurations
consisting of a monopole-antimonopole pair joined by a finite segment
of this putative string the expectation is that the configuration will
decay into a difuse and more-or-less spherical lump. The magnetic
charges will still be linked, as the string picture suggests, but the
confining character of the potential will be lost.

Of course it is possible that quantum corrections could modify our
results, but we have to stress that the stability of the solution
will, in general, depend on the {\it specific} details of the potential
\cite{H92}.  Neither is it sufficient to prove that the configurations
saturate a Bogomol'nyi bound. While BPS states are indeed stable,
saturation of a BPS bound does not preclude the existence of zero
modes or flat directions, which sometimes may result in an expansion
of the core of the string \cite{H92, GORS92}. Whether this can be
considered to lead to confinement is, at best, open to discussion.

\vspace{0.5truecm}
\noindent{\bf Acknowledgements}
\vspace{0.5truecm}

We are very grateful to Cumrun Vafa for thorough discussions, and to
I. Egusquiza, J.L.Ma\~nes and T. Vachaspati for their comments and suggestions.
A.~A.~acknowledges UPV grant 063.310-EB225/95 and CICYT grant
AEN96-1668.  This work is also supported by the European Commission
TMR programme ERBFMRX-CT96-0045, in which A.~A.~and M.~de R.~are
associated to the University of Utrecht.

\end{document}